# Solid-State Optical Magnetometer: Next-Generation Approach to Sub-Nanotesla Magnetic Sensing

O. Daneshmandi[1], M. Alidadi, Y.M. Banad[1], S. S. Sharif [1,2,*]

[1]School of Electrical and Computer Engineering, University of Oklahoma, Norman, OK
[2]Center for Quantum Research and Technology, University of Oklahoma, Norman, OK
[*] Corresponding Author

## ABSTRACT

This study presents a novel Solid-State Optical Magnetometer (SOM) based on black phosphorus (BP) multilayers, offering a highly sensitive, compact, and scalable alternative to conventional atomic-based optically pumped magnetometers. By leveraging BP's intrinsic linear dichroism property within a metasurface cavity, the proposed SOM achieves sub-nanotesla high-precision sensitivity in magnetic field detection and vector sensing capability. The integration of BP multilayers enhances light-matter interactions, enabling tunable optical responses driven by Lorentz force-induced cavity deformations. Optimized metasurface unit cells further improve polarization-dependent absorption, enhancing detection sensitivity and enabling precise field measurements. Finite Element Method simulations confirm that the SOM exhibits high linear ($R^2$ > 0.999) and tunable dynamic range, along with adaptable sensitivity via applied current modulation. The results demonstrate that increasing the applied current enhances sensitivity, achieving detection thresholds as low as 31.25 pT at 200 µA, while lower currents extend the dynamic range to ±10 nT at 50 µA. This tunable performance allows for application-specific optimization, making the device suitable for biomagnetic sensing (MEG, MCG), precision metrology, and industrial field detection. This work bridges the critical gap between atomic-based magnetometers and solid-state sensors by combining the high sensitivity of atomic systems with the miniaturization advantages of solid-state devices. Unlike SQUIDs requiring cryogenic cooling or OPMs constrained to millimeter dimensions by vapor cells, our BP-based SOM operates at room temperature with nanoscale dimensions while maintaining comparable sensitivity. The approach eliminates the need for alkali atoms in glass cells by replacing them with BP's anisotropic optical properties, while consuming less than 1 µW of power—orders of magnitude lower than conventional technologies. By establishing black phosphorus metasurface integration as a powerful tool for nanophotonic magnetic field detection, this work presents a low-power, miniaturized, and highly adaptable platform for next-generation magnetic field sensing technologies.

## I. INTRODUCTION

Magnetic field sensing has become integral to numerous scientific and research applications, from medical treatments and neurology studies to materials science and space exploration. The detection of sub-microtesla (µT) magnetic field variations presents significant challenges, requiring highly sensitive magnetometers for precise measurements across various domains.[1-3]

Current state-of-the-art magnetic sensing technologies include Superconducting Quantum Interference Devices (SQUIDs), atom-based Optically Pumped Magnetometers (OPMs)—particularly Spin-Exchange Relaxation-Free (SERF) OPMs—and advanced fluxgate magnetometers.[2, 4-9] These devices have demonstrated the ability to measure magnetic fields as low as 10 nanoteslas (nT), achieving sub-nanotesla sensitivity. However, each technology presents distinct limitations: SQUID-based magnetometers require complex cryogenic cooling systems, making them bulky, power-intensive, and expensive; OPMs, while demonstrating high sensitivity[5] since their breakthrough by the Romalis research group in 2002, they have faced challenges in miniaturization its dimensions due to their reliance on atomic vapor media, and fluxgate magnetometers, though offering vector detection capability, generally exhibit lower sensitivity compared to SQUIDs and OPMs.[8, 10-12]

To address these limitations, researchers have explored alternative approaches based on optical phenomena. These include fiber-optic magnetometers utilizing magneto-optical materials[12-14], Fabry-Pérot optical resonance systems[15], and metamaterial-based optical cavities.[16] For instance, Tang et al.[15] successfully integrated Fabry-Pérot interferometry with fiber-based structures, achieving external magnetic field detection in the milli- and sub-millitesla (mT) ranges with high linearity but limited sensitivity. Similarly, Lan et al.[17] introduced a metamaterial-based magnetometer at the sub-millimeter scale, demonstrating sub-microtesla sensitivity within a dynamic range that covers both mT and sub-mT fields. While these approaches have advanced miniaturization beyond traditional atomic-based OPMs, they still face challenges in achieving the sensitivity required for sub-nT detection[18].

Metasurface structures have emerged as a promising platform for further sensor miniaturization without sacrificing sensitivity.[19] These engineered surfaces excel in modulating phase, controlling polarization, shaping optical beams, and providing precise electromagnetic tuning.[20,21] The integration of metasurfaces with two-dimensional (2D) materials offers additional advantages, including enhanced light-matter interactions and tunable refractive indices for optimal light absorption. However, existing metasurface-based magnetometers require further improvement in sensitivity and tunability to compete with atomic-based systems.[21, 22]

Among 2D materials, black phosphorus (BP)[23] presents unique advantages for magnetometry applications that distinguish it from other 2D materials, such as graphene and transition metal dichalcogenides (TMDs). Unlike graphene, which lacks a bandgap, and TMDs, which exhibit isotropic optical properties, BP features intrinsic anisotropic electrical and optical responses due to its puckered lattice structure, a direct bandgap that is highly tunable (0.3 eV in bulk to 1.5 eV in monolayer form), strong optical nonlinearity with significant excitonic effects, and linear dichroism (LD) where a material absorbs light differently depending on its polarization direction. These properties make BP particularly suitable for integration into nanoscale metasurface structures for magnetic field detection.[24-26] Recent research has demonstrated that BP-based metasurfaces can function as highly tunable sensors, particularly for biosensing applications, where sensitivity can be adjusted by controlling BP layer thickness and metasurface geometry[27-33]. The strong light absorption and tunable bandgap of BP enable improved detection limits, faster response times, and enhanced sensing efficiency in photonic applications.[21, 34] Despite these promising capabilities, there is currently no established study demonstrating the use of BP's nano-optical phenomena for sub-μT magnetic field detection, particularly in the tens of nT range.

In this research, we introduce a novel Solid-State Optical Magnetometer (SOM) based on a black phosphorus multilayer metasurface structure, which operates without relying on traditional atomic-based Zeeman splitting mechanisms.[35, 36] Instead, our proposed SOM leverages BP's unique linear dichroism effect, combined with Lorentz force-induced cavity deformations, to achieve high-precision sub-nanotesla magnetic field detection. This approach addresses the fundamental limitations of existing magnetometry technologies by enabling:

- Ultra-compact device dimensions at the nanoscale
- Operation at room temperature without cryogenic cooling
- Vector detection capability for both magnitude and direction measurement
- Sub-nanotesla sensitivity with tunable dynamic range
- Low power consumption (≤1µW) suitable for portable applications

Table 1 provides a comparative overview of various state-of-the-art magnetic sensors, highlighting their key performance metrics, advantages, and limitations. This comparison underscores the need for a solid-state alternative, such as the proposed BP-based SOM, which offers high sensitivity, vector detection, and miniaturization while operating at ultra-low power.

The paper is structured as follows: Section I introduces the first-ever BP-based SOM, demonstrating its capability to detect magnetic field variations below tens of nanotesla with sub-nanotesla sensitivity. Section II presents comprehensive simulation results, highlighting the system's high linearity ($R^2 > 0.999$), dynamic range adaptability as a function of the applied current, and tunable sensitivity. These findings position BP and metasurface integration as a transformative approach for next-generation, high-sensitivity magnetometry, paving the way for wearable, low-power, and miniaturized magnetic sensing applications.

## I. DESIGN, SIMULATION DOMAIN, AND CHARACTERISTICS ANALYSIS OF THE BP-Based SOM

### A. Structural Design and Sensing Mechanism

The proposed Solid-State Optical Magnetometer is built upon a three-layer metasurface cavity, consisting of a top gold (Au) layer, a BP multilayer, and a bottom Au reflective layer. This layered structure is engineered to enhance light-matter interactions and enable high-sensitivity magnetic field detection through BP's linear dichroism effect and Lorentz force-induced cavity deformations (Figure 1(a)).

The top thin Au layer serves as the primary interface for the incident optical beam, facilitating plasmonic excitation and efficient light coupling into the metasurface cavity based on its CI-shaped unit cells, as shown in Figure 1(b). The plasmon excitation and the air cavity within the metasurface play essential roles in optimizing the SOM's performance. This enhancement of the electromagnetic wave confinement in the air cavity improves sensor sensitivity and facilitates constructive and destructive interference, ensuring optimal absorption efficiency. The middle BP multilayer is the defining feature of this SOM, providing an anisotropic optical response crucial for detecting external magnetic fields.

The air cavity gap, strategically positioned beneath the BP multilayer, serves two critical purposes: isolating surface plasmon resonances and optimizing the light coupling mechanism between the upper and lower Au layers based on Fabry-Perot resonance.

Table 1: Comparison of state-of-the-art magnetic field sensors, highlighting key performance metrics such as sensitivity, dynamic range, scalability, power consumption, and vector sensing capability. The proposed BP-based SOM offers a balance of high sensitivity, miniaturization, and tunability, addressing limitations of conventional SQUIDs, atomic OPMs, fluxgate magnetometers, and fiber-optic magnetometers while enabling low-power, scalable, and vectorial magnetic field detection.

| MAGNETOMETER TYPE | Dynamic Range | Sensitivity | Vector Capability | Power Consumption | Size & Scalability | Cryogenic Requirement | Applications |
|---|---|---|---|---|---|---|---|
| SQUID[7] | pT-μT | ~ fT/√Hz | No | in order of 1-10 W | Bulky, Requires Shielding | Yes | Neuroscience Research, Geophysics |
| ATOMIC OPMS[6, 9] | pT-μT | <pT/√Hz | No | in order of 1-10 W | Moderate (mm-scale cells) | No | Neuroscience Research, Biomedical Sensing, Geophysics, Geo-Space |
| FLUXGATE[4] | nT-μT | ~nT/√Hz | Yes | in order of <1 W | Bulky | No | Geophysics, Geo-Space |
| FIBER-OPTIC[15, 18] | mT-T | ~μT/√Hz | No | in order of mW | Ultra compact (nm-scale cells) | No | Industrial diagnoses |
| OPTICAL AND METAMATERIAL BASED[17] | μT-mT | ~nT/√Hz | Yes | in order of μW | Compact (μm-scale cells) | No | Neuroscience Research, Geophysics Biomedical Sensing and Industrial Diagnoses |
| PROPOSED BP-BASED SOM | pT-nT | %/ pT ≈ pT/√Hz | Yes | ≤1μW | Ultra compact (nm-scale cells) | No | Neuroscience Research, Geophysics Biomedical Sensing and Industrial Diagnoses |

This gap ensures the efficient confinement of electromagnetic waves, further boosting the system's sensitivity. The cavity length reference ($g_0$) was determined through parametric analysis, ensuring optimal performance which achieved under zero-field conditions.

The bottom Au layer functions as a reflective mirror, amplifying electromagnetic wave interactions within the cavity and BP multilayer. By reflecting light back into the structure, it dramatically enhances absorption while minimizing reflection. This configuration enforces the condition where the transmission spectrum approaches zero, establishing the relationship of absorption equal to one minus reflection.

The SOM sensing mechanism operates through a cascade of physical processes initiated by the Lorentz force. When an electric current (***I***) flows through the top Au layer in the presence of an external magnetic field (**B**), the resulting Lorentz force (***I***×**B**) generates mechanical displacement

($D_B$) in the top Au layer. This displacement directly modifies the cavity length according to $g=g_0 \pm D_B^{\pm}$, where the direction of displacement depends on the relative orientation of current and magnetic field vectors, (Figure 1(c)). This cavity deformation fundamentally alters the resonance conditions within the metasurface structure, which in turn modifies how BP interacts with incident light.

The modified cavity length changes the standing wave pattern within the Fabry-Pérot resonator, shifting the wavelength at which constructive interference occurs. This shift directly impacts BP's polarization-dependent absorption characteristics. Since BP exhibits strong linear dichroism, where s-polarized and p-polarized light are absorbed differently, any change in the cavity resonance conditions results in measurable changes in the absorption contrast between these polarization states. This differential absorption (ΔLD) serves as the primary sensing parameter, providing a direct optical readout of the external magnetic field's magnitude and direction.

To deeper understand this displacement for optimizing the sensitivity and dynamic range of the SOM, $D_B$ is governed by Lorentz Force ($I \times B$) and also the structural parameters of the metasurface. Thus, $D_B$ is expressed as[17]:

$$D_B \approx L^4 W^{-1} H^{-3}\ I \times B \qquad (1)$$

where L, W, and H define the metasurface's geometrical characteristics. This equation clarifies the variation of $D_B$ is equivalent to the magnetic field's changing while other parameters are fixed.

The BP multilayer's LD effect, where optical absorption varies with polarization, plays a critical role in sensing. The difference in absorption between s- and p-polarized incident light defines the LD, which serves as a critical concept for optical response of the BP-based SOM and its magnetic field detection. Based on LD property of BP multilayer and the cavity length variation which generated by Lorentz force in the SOM structure, as shown in Figure 1c, the external magnetic field can be sensed. The measurement of the absorbed light intensity according to its polarizations provides a spectrum that correlates with the existence of the external magnetic field, establishing a robust sensing mechanism. When no external field is present ($g=g_0$), the difference in absorption spectra between s- and p-polarized incident light defines the reference LD contrast ($\Delta LD_0$).[34] The presence of the external field results in the modified LD response for $g=g_0 \pm D_B^{\pm}$, and when reduced by $\Delta LD_0$, is named ΔLD.

$$\Delta LD_0 = A^s_{g0} - A^p_{g0} \qquad (2)$$
$$\Delta LD = (A^s_g - A^p_g) - \Delta LD_{g0} \qquad (3)$$

In these equations, $A^s_{g0}, A^p_{g0}$ and $A^s_g$ $A^p_g$ are the absorption values when the illuminated beam polarization is s and p, and $g_0$ and $g$ values for cavity length, respectively. The ΔLD parameter serves as the figure of merit (FOM) for magnetic field detection by the BP-based SOM.

The synergy between the Au thin films, the BP multilayer, and the cavity gap results in a meticulously engineered plasmonic design that exploits both plasmonic resonance and BP's LD effect to achieve ultra-sensitive magnetic field detection. Together, these elements create a scalable and high performance platform for next-generation nanoscale magnetometry, offering tunable dynamic ranges and sub-$nT$ sensitivities. Further details on the parametric selection of the metasurface structure, the cavity length, and its impact on performance, as well as the simulation setup, are provided in Section B.

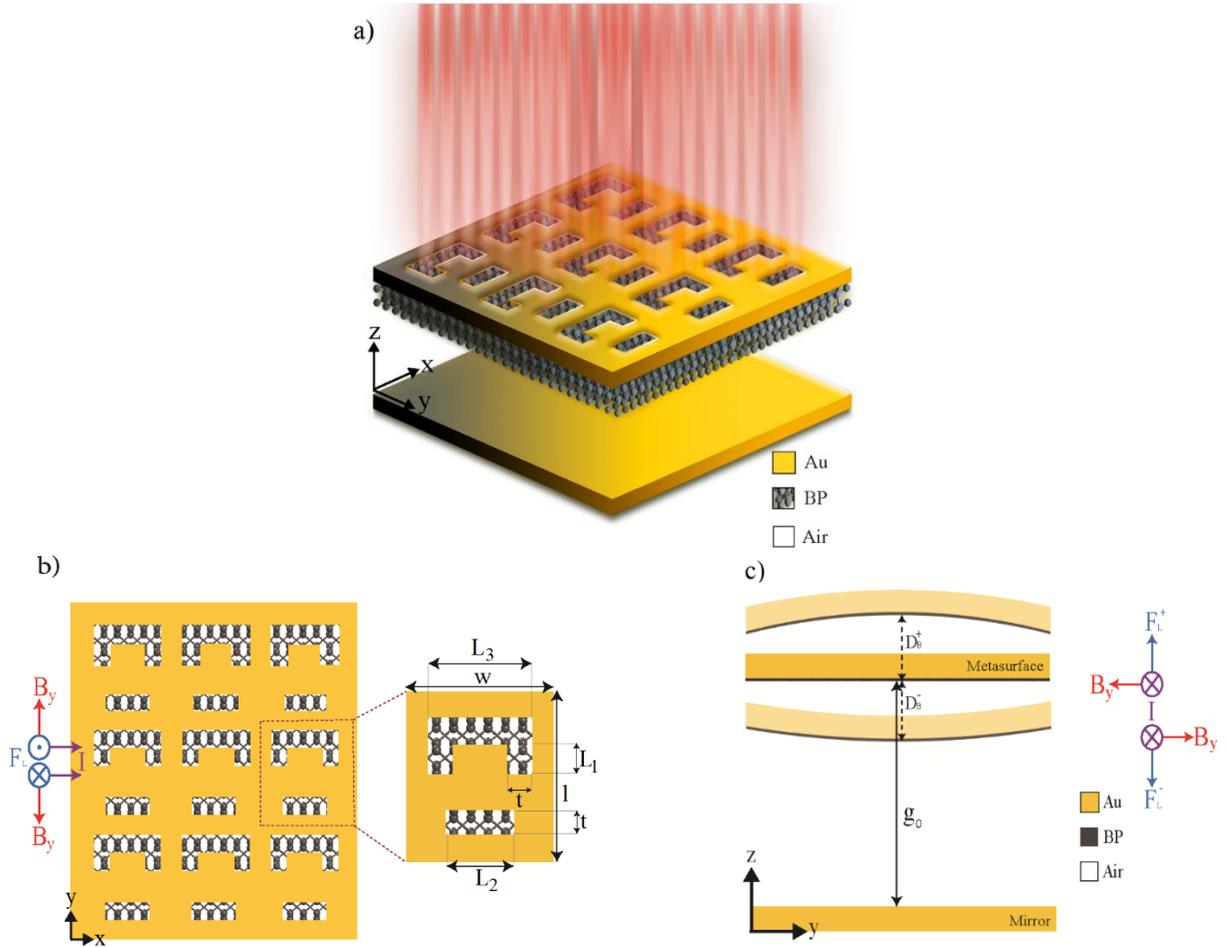

Figure 1: Structural design of the BP-based Solid-State Optical Magnetometer. (a) Schematic showing the three-layer structure with top gold layer (50 nm), black phosphorus multilayer (2 nm), and bottom gold reflective layer (50 nm) separated by an air cavity of length $g_0$ = 900 nm. The xyz coordinate system indicates the orientation of the device. (b) Top view of the metasurface CI-patterned unit cell with key dimensions: W = 400 nm, l = 450 nm, t = 75 nm, $L_3$ = 300 nm, $L_1$ = 60 nm, and $L_2$ = 300 nm. Applied current flows through the top gold layer in the x-direction, while the external magnetic field can be applied in the y-direction, enabling vector sensing capability. (c) Cross-sectional view showing how Lorentz force ($I \times B$) induces displacement ($D_B$) in the top gold layer, altering the cavity length to g = $g_0 \pm D_B$ and modifying optical absorption characteristics.

## B. Computational Modeling and Simulation Setup

To accurately model the performance of the proposed BP-based SOM, the FEM was employed due to its effectiveness in solving complex electromagnetic problems involving intricate geometries, heterogeneous materials, and boundary conditions. FEM provides a robust framework for simulating electromagnetic field propagation within the metasurface-integrated structure. The mesh resolution was set to 1/20 of the minimum wavelength to ensure numerical accuracy, while periodic boundary conditions were applied along the x- and y-axes. To provide artificial reflections and simulate an open domain, a Perfect Match Layer (PML) was implemented along the z-axis.

The metasurface cavity is further structured with 3×3 CI-shaped unit cells, as shown in Figure 1(b), where the unit cell dimensions (W = 400 nm, l = 450 nm, t = 75 nm, $L_3$ = 300 nm, $L_1$ = 60 nm, and $L_2$ = 300 nm) are optimized to enhance light-matter interactions. This structured design ensures efficient polarization-dependent absorption, maximizing BP's LD effect while maintaining plasmonic resonance within the cavity. The cavity reference length ($g_0$ = 900 nm) was determined through parametric analysis for optimal performance under zero-field conditions, with the metasurface cavity functioning as a Fabry-Pérot resonator. This parameter is one of several facilities available to ensure the optimized absorption of incident light by the proposed BP-based SOM.

The optical properties of the gold layers, functioning as both the metasurface and the reflective mirror, were modeled using the Drude-Lorentz approach, with each layer set to a 50 *nm* thickness. The BP multilayer, precisely engineered at 2 *nm*, was incorporated as a semiconductor within a dielectric-metal-semiconductor (Air-Au-BP) configuration, where its linear dichroism effect plays a crucial role in the sensing mechanism. The Drude model was employed to characterize BP's optical response with particular emphasis on the differences in effective electron masses along the armchair and zigzag directions, which govern its anisotropic absorption properties.[37-39] Further details on BP's optical and electronic properties, including the equations for conductivity and permittivity matrices, are provided in the Supplemental Materials (S.1). These simulations establish the SOM's capacity for highly sensitive, tunable magnetic field detection, forming the foundation for subsequent performance analysis.

## II. RESULTS AND DISCUSSION

### A. Absorption Spectra Analysis and Magnetic Field Detectivity

The magnetic sensing behavior of the designed SOM, driven by the LD property of the BP multilayer, was first analyzed by investigating its absorption spectra when illuminated by a plane wave optical beam, as described in Figure 2. The absorption spectra for both s- and p-polarized beams were analyzed when the cavity length was set to $g_0$ with no magnetic field. Due to the inclusion of a thin Au film functioning as a mirror, the structure operates in the reflection regime, resulting in negligible transmission. The results indicate that maximum absorption occurs at a wavelength of 980 *nm*. This peak absorption wavelength results from the interplay between plasmonic excitation and Fabry-Pérot resonance within the metasurface cavity along with BP's intrinsic LD effect, which is highly sensitive to light polarization. This wavelength was selected as a reference point for comparing the device's absorption performance under the two s- and p-polarization states, according to Equation (2).

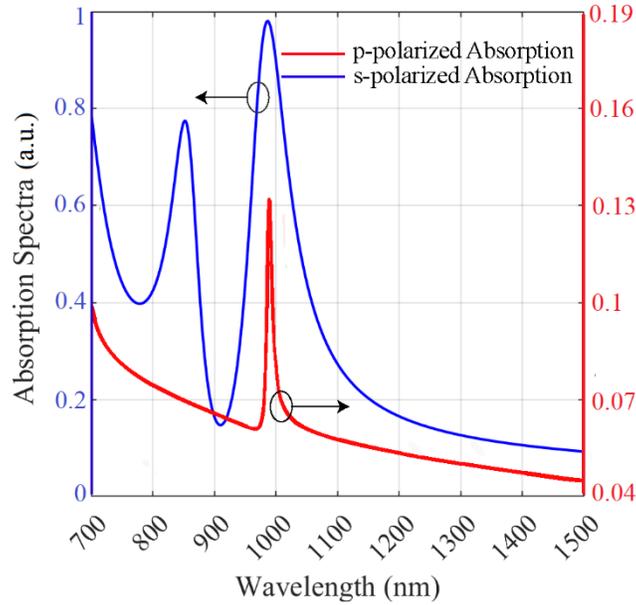

Figure 2: Absorption spectra of the initial BP-based SOM design under zero magnetic field (B = 0) conditions. The graph shows distinctly different absorption for s-polarized (blue line) and p-polarized (red line) incident light due to BP's linear dichroism property, with maximum s-polarized absorption of 98.07% and p-polarized absorption of 13.46% occurring at λ = 980 nm. This polarization-dependent absorption difference serves as the baseline for magnetic field detection using the linear dichroism contrast (ΔLD) parameter.

## B. Optimization of Metasurface Unit Cell for Enhancing Magnetic Detectivity

To maximize the magnetic field detectivity of the proposed SOM, we optimized the unit cell geometry while analyzing the physical mechanisms behind the polarization-dependent absorption that defines the linear dichroism contrast. Since the SOM's sensing mechanism relies on BP's LD effect, enhancing the contrast between s- and p-polarized absorption spectra directly improves sensitivity.

The optimization process focused on tuning the structural parameters of the unit cell at steady states ($g_0$, **B**-field=0) to achieve maximum light-matter interaction at a wavelength where the metasurface exhibits strong resonance leading to maximum absorption. The cavity length was chosen through a detailed optimization process involving parametric sweeps across various lengths. This analysis examined how different cavity lengths influence the constructive and destructive interference patterns within the Fabry-Pérot resonator. Various configurations were simulated with cavity lengths ranging from 700 nm to 1100 nm to determine the optimal resonance conditions that maximize absorption in the initial structure of the designed SOM. The reference cavity gap, was defined under a non-magnetic field condition and was established based on multiple simulations, with results presented in the Supplemental Materials (S.2). Ultimately, at a length of 900 nm, the standing wave pattern aligns optimally, resulting in minimal reflection and maximum absorption of the SOM, which is crucial for subsequent steps in the SOM design process.

Furthermore, the geometrical parameters $L_1$ and $L_2$ demonstrated a significant impact on the absorption properties due to their mutual coupling effect, directly influencing the sensitivity of the BP-based SOM. We systematically varied these parameters through numerical simulations to identify an optimal configuration that maximizes absorption contrast. Figure 3 shows the results of varying $L_1$ from 45 to 105 nm and $L_2$ from 200 to 350 nm. Multiple configurations within these ranges yield high absorption, with an optimized performance at $\lambda = 960$ nm. For our simulation purposes, we selected $L_1 = 60$ nm and $L_2 = 200$ nm, as this configuration represents one of the optimal points within the high-absorption region. This tolerance in design parameters is particularly advantageous for practical implementations, as it ensures robustness against small variations in fabrication.

In the optimization process of the proposed SOM, our primary goal was to maximize its sensitivity. We focused on achieving optimal values for the relevant parameters to meet this objective. Additionally, based on the feasibility study of near-infrared (NIR) wavelengths for the incident wave passing through the nanostructure—which is advantageous for optical integrated circuits—we ensured that the wavelength corresponding to the maximum absorption remained within the range of 900 to 1100 nm.[40]

Figure 4 presents the optimized absorption spectra of the SOM structure, illustrating the enhancement achieved through unit cell optimization. Compared to the initial absorption spectra in Figure 2, a notable increase in absorption is observed for s-polarized light, improving from 98.07% to 99.95% at the maximum absorption wavelength. Conversely, the absorption of p-polarized light decreases significantly from 13.46% to 9.76%. This behavior is explained by both the enhanced localized surface plasmon resonance (LSPR) and BP's anisotropic absorption response.

This polarization-dependent absorption arises from the distinct field distributions within the metasurface cavity, as illustrated in Figure 5. For s-polarized light, the electric field aligns with BP's high-absorption armchair direction, creating stronger plasmonic field confinement and maximizing interaction with the BP layer. In contrast, p-polarized light interacts with BP's low-absorption zigzag direction, resulting in weaker energy coupling and lower overall absorption. The metasurface cavity further enhances this contrast by functioning as a Fabry-Pérot resonator that amplifies the difference between polarization states.

These differences in field distributions and absorption behaviors are critical for the SOM's magnetic sensing mechanism. When an external field is applied, the Lorentz force alters the cavity gap ($g=g_0 \pm D_B$), leading to real-time modulation of the LD contrast. The plasmonic field confinement ensures that even minor deformations in the cavity result in measurable optical shifts, making the SOM highly sensitive to weak magnetic fields. The optimized design not only enhances overall absorption efficiency but also strategically modifies polarization-dependent absorption, leading to a stronger, more tunable magneto-optical response.

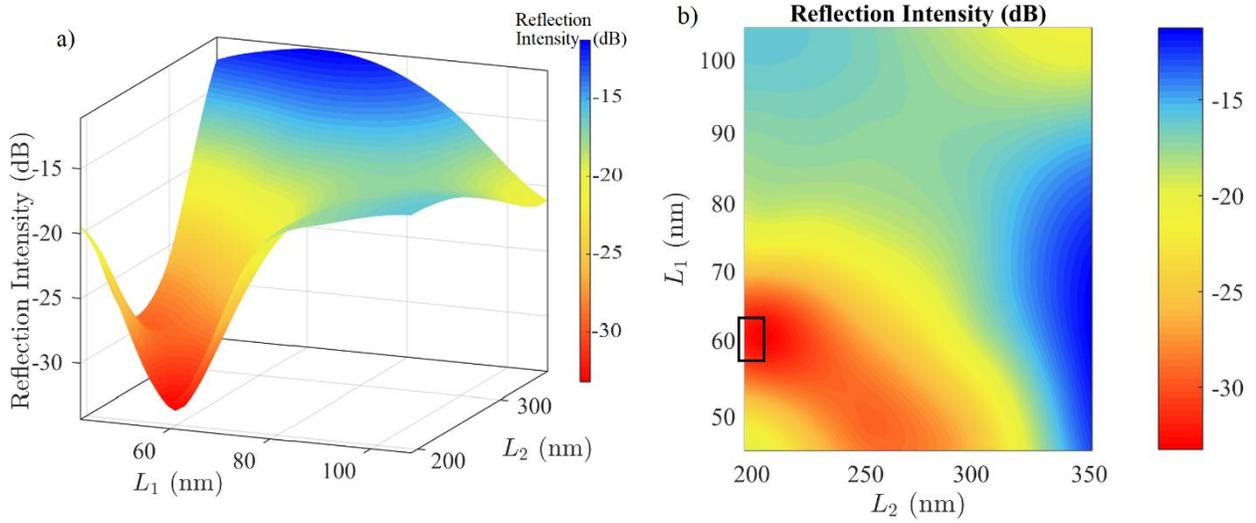

Figure 3: Optimization map for the metasurface unit cell dimensions $L_1$ and $L_2$. The color scale represents the minimum reflection intensity of the SOM structure by varying the $L_1$ and $L_2$ values. The blue regions indicate high reflection, while the red regions signify low reflection. The optimization study covered $L_1$ values from 45-105 nm and $L_2$ values from 200-350 nm, with optimal performance (minimum reflection/maximum absorption) occurring around $L_1$ = 60 nm and $L_2$ = 200 nm (rectangular marked region). The broad rectangular region demonstrates fabrication tolerance, where small variations in dimensions still maintain high absorption performance.

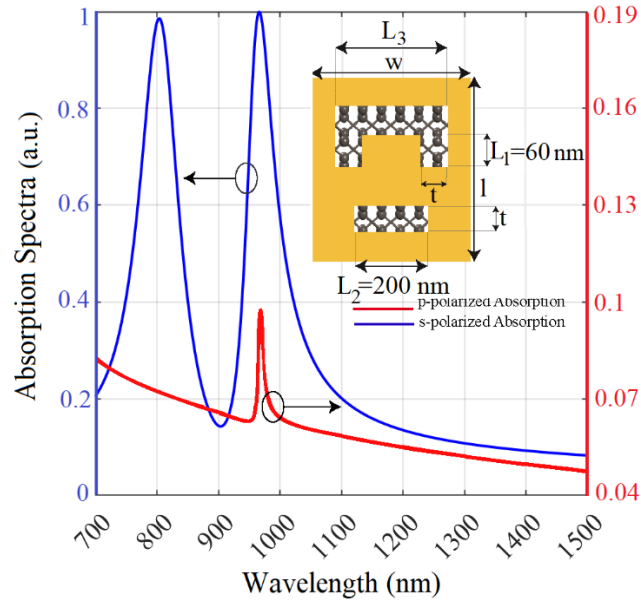

Figure 4: Absorption spectra of the optimized BP-based SOM design with unit cell parameters $L_1$ = 60 nm, $L_2$ = 200 nm, W = 400 nm, l = 450 nm, t = 75 nm, and $L_3$ = 300 nm at cavity length $g = g_0$. The optimized design shows significantly enhanced polarization contrast compared to Figure 2, with s-polarized absorption (blue line) reaching 99.95% and p-polarized absorption (red line) reduced to 9.76% at $\lambda$ = 960 nm. This increased contrast directly improves the sensitivity of the SOM by enhancing the linear dichroism ($\Delta$LD) parameter used for the magnetic field detection mechanism.

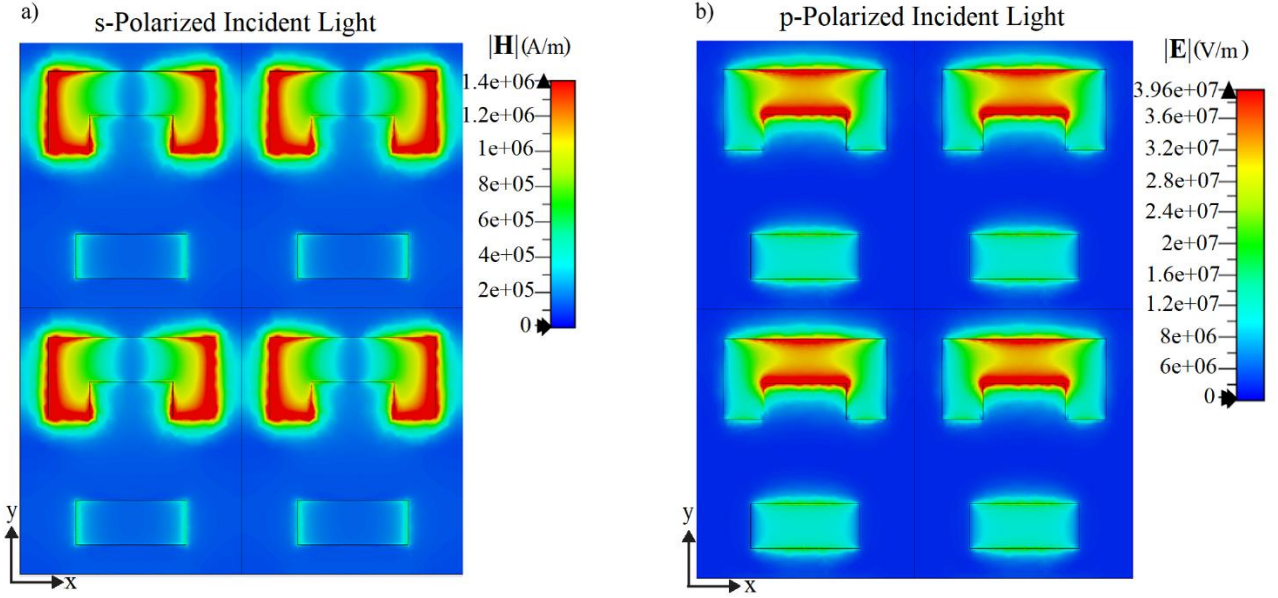

Figure 5: Electromagnetic field distribution in the optimized SOM structure at λ = 960 nm and zero magnetic field. (a) Distribution of magnetic field intensity norm for s-polarized incident light, showing strong field confinement within the metasurface cavity. (b) Electric field intensity norm distribution for p-polarized incident light, demonstrating weaker field penetration into the BP layer. These different field distributions explain the large absorption contrast between polarization states, with s-polarized light aligning with BP's high-absorption armchair direction and p-polarized light aligning with BP's low-absorption zigzag direction.

## C. Correlation of Linear Dichroism and Tunable Magnetic Field Detectivity

To confirm the linear dichroism property and evaluate the detectivity and performance of the designed SOM, we analyzed the ΔLD spectra, which represent the difference in absorption between s- and p-polarized light under varying external magnetic fields. These spectra were measured based on several magnitudes of the flowed current through the x-axis while applying different magnetic field strengths along the y-axis, as illustrated in Figure 1. Since the cavity length (g) is influenced by Lorentz force-induced displacement ($D_B$), which depends on both **B** and **I**, any changes in the **B**-field modify the absorption characteristics of the metasurface. As a result, the ΔLD spectra dynamically shift in response to the external magnetic field, making ΔLD a key FOM for SOM detectivity. Figure 6 presents the ΔLD spectra for different applied current values (±50, ±100, and ±200 μA), demonstrating how the SOM's absorption response evolves under varying external magnetic fields. The tested magnetic field strengths range from 125 *pT* to 5 *nT*, covering weak field detection scenarios relevant for ultra-sensitive magnetometry applications. The applied current flowing through the metasurface layer must remain within the tolerable current density limit of the Au thin film ($10^5$ *A/cm²*) to prevent structural or performance degradation.[41]

As shown in Figure 6, the results demonstrate that at any given magnetic field strength, the ΔLD spectra vary proportionally with the applied current. The increase in B-field magnitude leads to a corresponding rise in ΔLD, with the most significant changes observed at λ = 960 nm, where the metasurface exhibits maximum absorption due to plasmonic resonance enhancement. A detailed examination of each subfigure reveals important patterns that validate the SOM's capabilities:

In Figure 6(a), which shows the response to a 125 pT field, the ΔLD's maximum value reaches approximately 0.17% at 50 µA, 0.32% at 100 µA, and 0.67% at 200 µA, demonstrating a near-linear scaling with current at this extremely low field strength. This confirms the device's sub-nanotesla sensitivity even at modest current levels, operating at room temperature without cryogenic cooling requirements that burden SQUID-based systems. The symmetrical response to positive and negative currents (-50, -100, and -200 µA) producing nearly identical magnitude but opposite-sign ΔLD values validates the vectorial detection capability, allowing determination of both field magnitude and direction.

Comparing Figures 6(a) through 6(c), the ΔLD response in the picotesla range (125-500 pT) shows exceptional linearity with increasing field strength. At 200 µA, the peak ΔLD increases from 0.67% at 125 pT to 2.7% at 500 pT. This performance is achieved within the metasurface's nanoscale dimensions, representing a significant advance over conventional magnetometers that require millimeter to centimeter-scale components. In the nanotesla range (Figures 6(d) through 6(f)), the ΔLD continues to scale proportionally with field strength, maintaining a consistent relationship between current and response amplitude. At 5 nT (Figure 6(f)), the peak ΔLD reaches 27.5% at 200 µA, representing a 33-fold increase from the 0.67% observed at 125 pT with the same current. This wide dynamic range combined with high sensitivity is crucial for applications ranging from biomagnetic sensing to industrial diagnostics.

The power consumption remains extremely low across all measurement scenarios. Even at the highest current (200 µA) flowing through the thin gold layer with resistance in the order of 1-10 Ω, the power consumption remains below 1 µW. This ultra-low power operation is orders of magnitude below conventional magnetometers, enabling battery-powered operation for portable and wearable applications.

A particularly significant observation is how the spectral shape evolves with increasing field strength. At lower field strengths (Figures 6(a) and 6(b)), the ΔLD curves maintain a relatively symmetric profile around the resonance wavelength of 960 nm. However, as the field strength increases (Figures 6(e) and 6(f)), the spectra develop more pronounced asymmetry, with secondary features appearing at shorter wavelengths. This spectral evolution provides additional information about the field characteristics beyond simple amplitude measurements, potentially allowing for more sophisticated field analysis in practical applications.

The detection of a constant magnetic field is enhanced as the applied current increases. For example, Figure 6 shows that when the magnetic field strength is set to 125 pT, the change in optical density increases by approximately four times when the current shifts from 50 µA to 200 µA. This increase in ΔLD is also observed at all magnetic field strengths as shown in Figure 6, with a similar increase in the applied current. This significant improvement in ΔLD results from the increased current value in equation (1), which leads to a rise in $D_B$, subsequently affecting the absorption spectrum and ΔLD.

For a fixed applied current of I = 100 µA, increasing the external field from B = 125 pT to 500 pT results in a ΔLD shift of approximately 4 times (from 0.319 to 1.354), whereas in the nT range, increasing B magnitude from 1 nT to 5 nT produces a ΔLD change of approximately 5 times (from 2.735 to 13.794). These variations indicate that the SOM exhibits higher sensitivity in the low-field (pT) regime, making it particularly suitable for applications requiring ultra-precise magnetometry such as magnetoencephalography, where neural magnetic fields are typically in the pT to nT range.

The consistency of these responses across multiple field strengths and current values confirms that the BP-based SOM achieves all five key capabilities claimed in the introduction: (1) nanoscale dimensions through its metasurface design, (2) room-temperature operation without cryogenic requirements, (3) vector detection capability through bidirectional current and field response, (4) sub-nanotesla sensitivity with tunable dynamic range via current modulation, and (5) low power consumption under 1 µW suitable for portable applications. These capabilities position the BP-based SOM as a transformative technology that bridges the gap between conventional atomic-based magnetometers and solid-state sensors.

As shown in Figure 6, the results demonstrate that at any given magnetic field strength, the ΔLD spectra vary proportionally with the applied current. The increase in **B**-field magnitude leads to a corresponding rise in ΔLD, with the most significant changes observed at λ = 960 nm, where the metasurface exhibits maximum absorption due to plasmonic resonance enhancement. This confirms that the optimized metasurface structure is highly responsive to external field perturbations, reinforcing the role of BP's anisotropic absorption in enhancing detectivity.

The detection of a constant magnetic field is enhanced as the applied current increases. For example, Figure 6 shows that when the magnetic field strength is set to 125 pT, the change in optical density (ΔLD) increases by approximately four times when the current shifts from 50 µA to 200 µA. This increase in ΔLD is also observed at all magnetic field strengths as shown in Figure 6, with a similar increase in the applied current. This significant improvement in ΔLD results from the increased current value in equation (1), which leads to a rise in $D_B$, subsequently affecting the absorption spectrum and ΔLD.

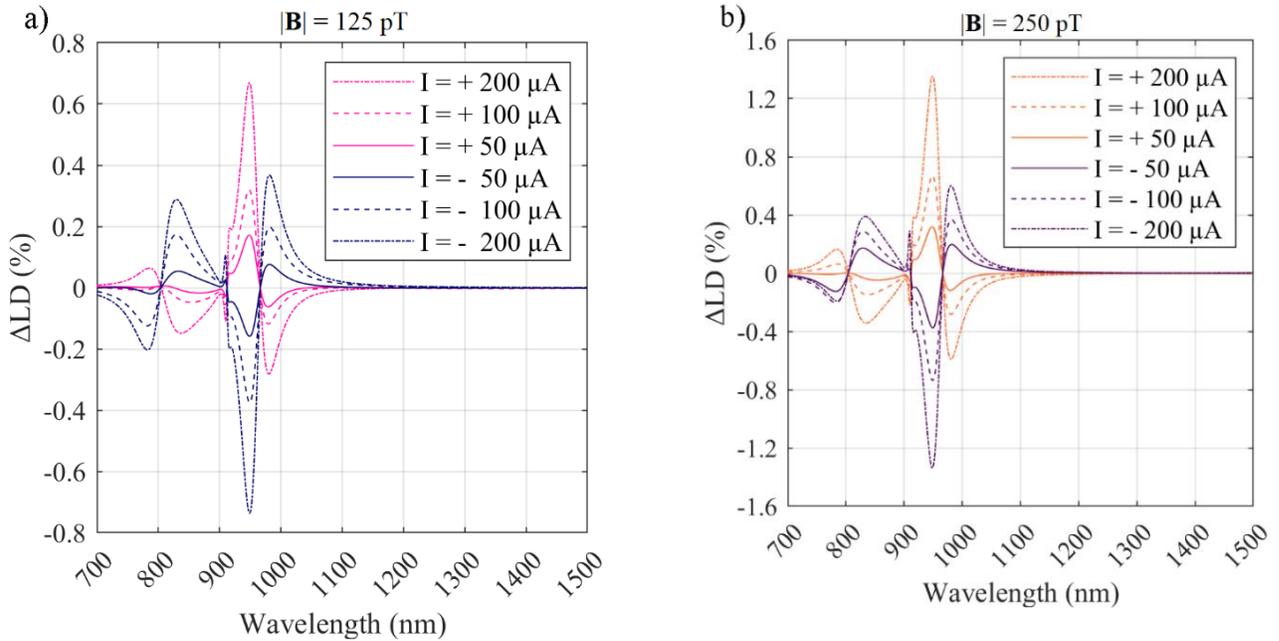

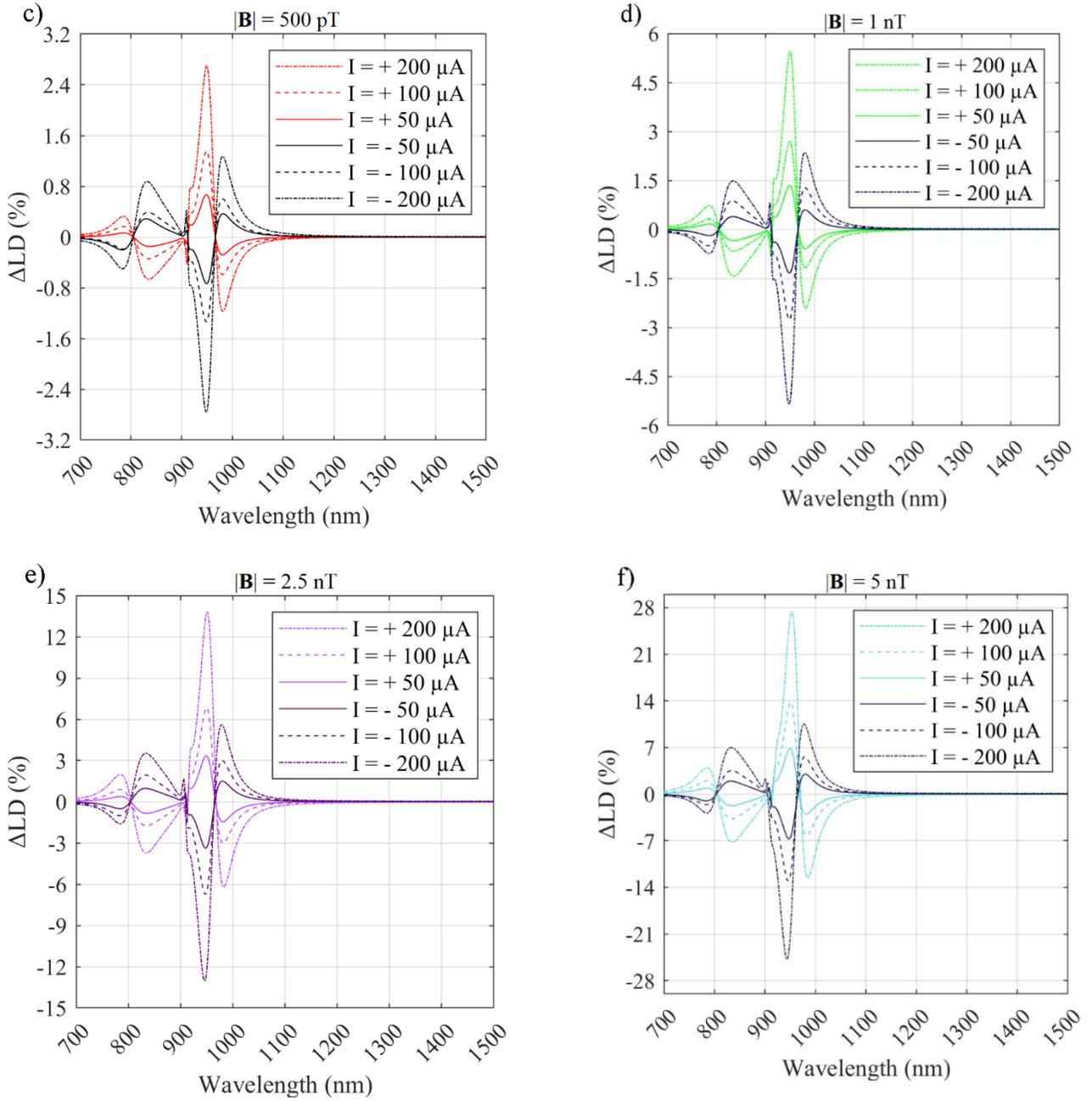

Figure 6: ΔLD spectra of the SOM under varying external magnetic field strengths and applied currents. Panels show responses to: (a) 125 pT, (b) 250 pT, (c) 500 pT, (d) 1 nT, (e) 2.5 nT, and (f) 5 nT magnetic fields. Each panel displays ΔLD for six different applied current conditions (±50, ±100, and ±200 µA), demonstrating how sensitivity scales with current magnitude and how sign reversal of either current or magnetic field inverts the ΔLD response, enabling vector field detection. Peak ΔLD values at λ = 960 nm range from 0.17% (50 µA, 125 pT) to 27.5% (200 µA, 5 nT), confirming sub-nanotesla sensitivity across a wide dynamic range.

## D. The Presented SOM Detectivity Characterization Based on ΔLD Versus Magnetic Field Strength

Based on the functional behavior of the ΔLD versus variation of exposed magnetic field magnitudes for the SOM, this magnetic sensor should be characterized according to several critical criteria that define its magnetic field detectivity. The most important properties of a magnetometer are linearity, dynamic range, and sensitivity, which in the case of the designed SOM, can be determined from the reference curve as its functional. Figure 7 illustrates the variation of ΔLD spectra as a function of external magnetic field strength (**B**), providing critical insights into the SOM's dynamic range, linearity, sensitivity, and vectorial detection capability. The ΔLD spectra for a fixed applied current (I = 50 µA) under different magnetic field magnitudes are shown in Figure 7(a), where the external field is directed along the y-axis of the SOM structure. The results indicate that when a consistent current flows through the metasurface layer (Figure 1(b)) and the magnetic field intensity increases, the ΔLD percentage increases at λ = 960 nm, where the metasurface exhibits maximum absorption contrast due to plasmonic resonance enhancement. Moreover, when the applied current remains constant, the ΔLD values become negative when the B-field direction is reversed, confirming the SOM's capability to distinguish magnetic field polarity and enabling vectorial magnetic field detection.

As shown in Figure 7(b), the ΔLD values at λ = 960 nm correlate strongly with the external magnetic field strength, demonstrating a linear response. We characterized the linearity using Ordinary Least Squares regression analysis, with the Coefficient of Determination ($R^2$) serving as the primary metric. A segment of the curve in Figure 7(b) was fitted with a high degree of linearity, achieving an $R^2$ value above 0.999. We define the dynamic range as the span of magnetic field strengths over which this high linearity ($R^2 > 0.999$) is maintained. Importantly, this dynamic range can be tuned by adjusting the current passing through the device. The ΔLD spectra were also analyzed for higher applied currents (100 µA and 200 µA), with results provided in the Supplemental Materials (S.3).

Figure 8 illustrates the variation of ΔLD at λ = 960 nm as a function of the external magnetic field strength for different applied currents (I = 50, 100, and 200 µA), when its linearity is $R^2 > 0.999$. The trends observed in Figure 8(a), corresponding to the initial SOM structure, and Figure 8(b), representing the optimized design, provide critical insights into the sensor's magnetic field detectivity, dynamic range, and sensitivity modulation. The results indicate that increasing the applied current enhances the sensitivity of the SOM, as evidenced by the steeper slopes in the ΔLD versus B curves. However, this increase in sensitivity comes at the cost of a reduced dynamic range based on the same linearity. For instance, when the applied current is set to 50 µA, the SOM exhibits a wider dynamic range of ±10 nT, whereas increasing the current to 200 µA reduces the dynamic range to ±2.5 nT. This behavior aligns with the Lorentz force-induced displacement model described in Equation (1), where **$D_B$** scales with **I×B**, leading to stronger optical modulation at higher currents.

Additionally, the sensitivity of the magnetometers and the presented SOM is inversely proportional to the slope of the magnetic field detection curve of a magnetometer and by enhancing its sensitivity the amount of a net magnetic field which it can detect, decreases ($\delta |B_y| \sim (\frac{d(\Delta LD)}{d(B_y)})^{-1}$).[42]

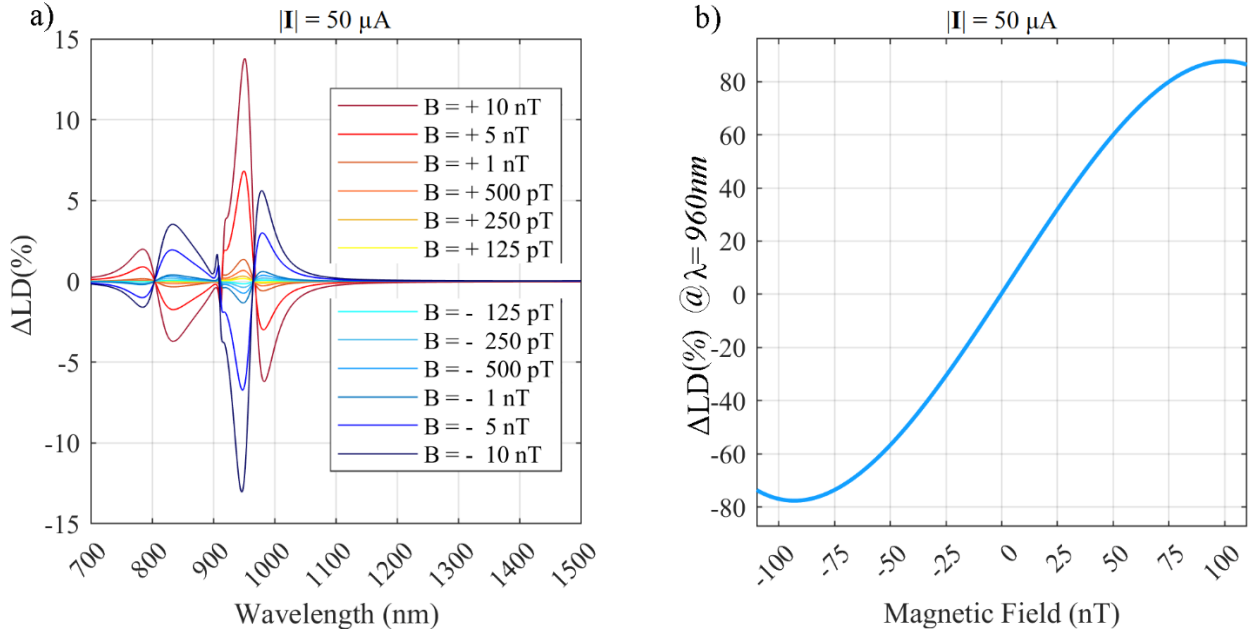

Figure 7: (a) ΔLD spectra of the SOM with 50 µA applied current under varying magnetic field strengths (-5 nT to +5 nT), showing proportional response to field intensity and sign inversion with field direction reversal. (b) The plot of ΔLD values at λ = 960 nm versus magnetic field strength demonstrates high linearity ($R^2 > 0.999$) across the ±5 nT range. The linear relationship between ΔLD and magnetic field strength enables precise field magnitude measurement, while the sign of ΔLD indicates field direction, confirming the device's vector sensing capability.

A comparison between Figure 8(a) and 8(b) highlights the impact of metasurface optimization due to selected factors, on the performance of the SOM. The optimized metasurface unit cell, particularly with structural parameters $L_1$=60 nm and $L_2$=200 nm, significantly improves the contrast between s- and p-polarized absorption, thereby increasing the overall ΔLD response. This indicates that the main objective of optimizing the SOM has been successfully achieved, resulting in improved sensitivity. The slope of the ΔLD versus magnetic field for each applied current increases, and more detailed comparisons for each current can be found in the Supplemental Materials (S.4).

Moreover, as demonstrated in Figure 8b, the rate of ΔLD variation per unit field strength depends on the applied current. When the current increases from I = 50 µA to 200 µA, the ΔLD response per unit field strength (ΔLD/nT) improves from 1.35 %/nT to 5.38 %/nT, demonstrating a tunable sensitivity mechanism where larger applied currents enhance the system's response to weak magnetic fields. This effect arises from the increased Lorentz force ($I \times B$) contribution, which amplifies the cavity length modulation ($D_B$) and consequently the ΔLD contrast. These findings confirm that by adjusting the applied current, the SOM can be optimized for specific sensitivity requirements, allowing for high-resolution detection of weak magnetic fields down to the sub-nT level. The ability to finely tune the ΔLD response by modulating current levels makes this platform highly versatile for next-generation precision magnetometry applications.

These variations indicate that the SOM exhibits higher sensitivity in the low-field (pT) regime, making it particularly suitable for ultra-precise magnetometry applications. Since the metasurface cavity functions as a Fabry-Pérot resonator, even small variations in $D_B$ induce measurable shifts in optical absorption. This behavior establishes $D_B$ as a key factor in determining the system's detectivity, with larger $I$ values enhancing sensitivity at the cost of reduced dynamic range.

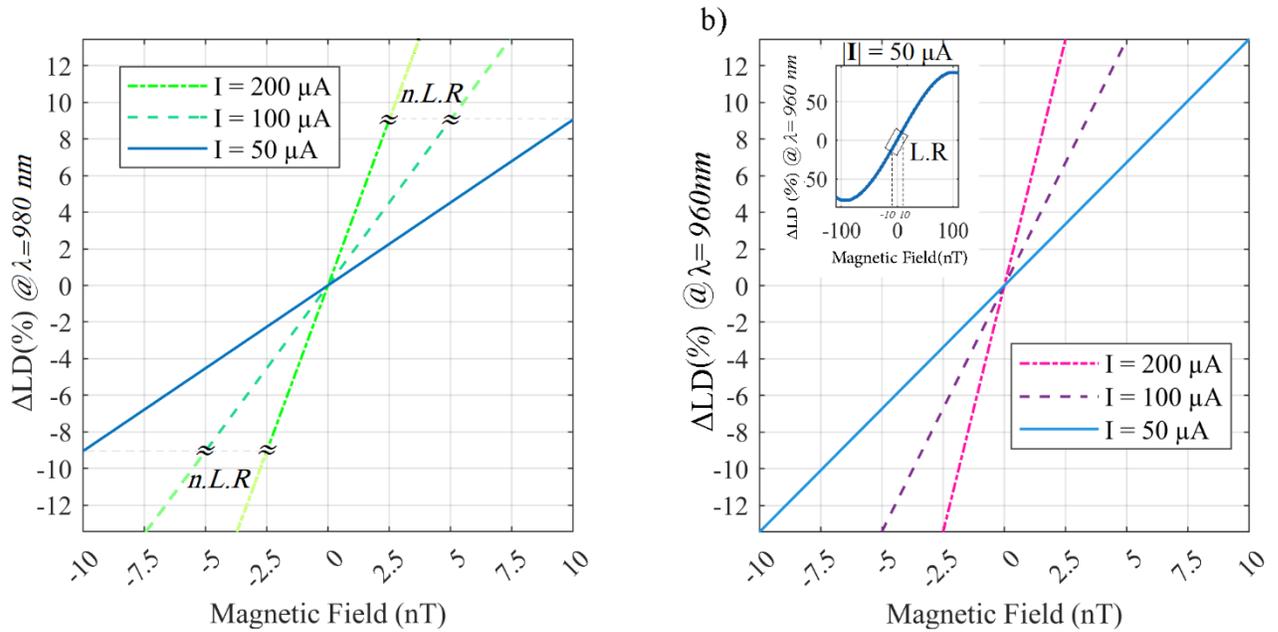

Figure 8: Magnetic field sensitivity characterization showing ΔLD at the the minimum reflection wavelengths versus magnetic field strength for different applied currents. (a) Performance of the initial SOM design which distincts non-linear region (n.L.R) of the initial structure and (b) performance of the optimized design with enhanced linear dichroism contrast in the linear region (L.R.) of the ΔLD versus strength of magnetic fields. Both graphs demonstrate how the optimization process improves the sensor sensitivity, and also increasing applied current from 50 μA to 200 μA enhances sensitivity (steeper slope) at the cost of reduced dynamic range. The optimized design achieves sensitivity values of 1.35 %/nT at 50 μA (±10 nT range), 2.70 %/nT at 100 μA (±5 nT range), and 5.38 %/nT at 200 μA (±2.5 nT range), enabling application-specific tuning between sensitivity and dynamic range.

Another key observation is that ΔLD changes polarity when the direction of B is reversed, indicating that the SOM is capable of vectorial magnetic field detection. This means the SOM can measure not only the magnitude of the external field but also its direction, which is a critical feature for applications such as biomagnetic sensing (e.g., MEG, MCG), neuromagnetic imaging, and spintronic devices. The ability to resolve field directionality further establishes the SOM as a versatile tool for nanoscale and quantum sensing applications.

These findings confirm that the SOM exhibits a linear, highly tunable response to external magnetic fields, with sub-nT detection sensitivity. The ability to adjust the detection range by varying the applied current makes it a highly flexible platform for ultra-sensitive, low-power magnetometry applications, particularly in wearable biomedical sensors, neuromagnetic monitoring, and quantum metrology.

## E. Tolerance Study: Impact of the Tolerance Fabrication on the Proposed SOM Properties

The proved concept of the designed BP-based SOM utilized FEM simulations that have been validated, leading to the proposal of a novel structure for magnetic field detection. However, practical implementation raises some concerns. As a nanostructure, manufacturing tolerances, the accuracy of the applied currents, and the inherent characteristics of BP as a 2D material significantly impact the SOM feature dimensions. According to Equation (1), all structural parameters of the SOM, which are related to manufacturing techniques and the precision of the currents flowing through the metasurface layer, influence variations in the DB value. Suppose a tolerance of ±5% for the $D_B$ magnitude is considered practically due to all these mentioned tolerances. In that case, this variance can be added to or subtracted from its values based on constant structural parameters and the applied current. These tolerance effects become particularly evident in the variations of ΔLD versus the strength of the $B$-field (Figure 8(b)), as it pertains to the magnetic field detection capabilities of the SOM.

When a current magnitude of 50 µA is applied, the detectivity behavior of the SOM is illustrated by the relationship between ΔLD and different strengths of the exposed magnetic field when the SOM is manufactured to fixed physical values and applied current intensity (Figure 9). In contrast to this reference linear region of the SOM's dynamic range, the ±5% tolerance in DB affects the slope of the ΔLD curve in response to the exposed magnetic field, indicating changes in the sensitivity of the SOM. As demonstrated in Figure 9, a +5% tolerance in the parameter DB results in an increase of approximately 4.9% (1.412 %/nT) in the slope of the ΔLD curve compared to the scenario without $D_B$ tolerance (1.345 %/nT).

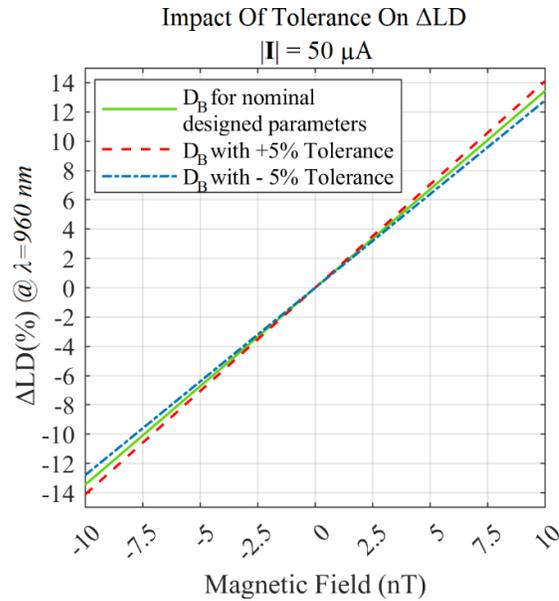

Figure 9: Impact of fabrication and current tolerance on the SOM's magnetic field detectivity. The graph shows ΔLD versus magnetic field strength for the optimized design with 50 µA applied current under three conditions: nominal design parameters (solid line, 1.345 %/nT sensitivity) and ±5% tolerance in the displacement parameter DB (dash and dash-dot lines, resulting in 1.412 %/nT and 1.28 %/nT sensitivity respectively). This analysis demonstrates that typical manufacturing variations and current precision limitations would affect sensor sensitivity by approximately ±5%, confirming the robustness of the design for practical implementation.

Conversely, a -5% tolerance in $D_B$ leads to a decrease of about -4.9% (1.28 %/nT) in the slope of the ΔLD variation concerning the magnitude of the magnetic field compared to the situation without $D_B$ tolerance. This shows that any tolerances arising from fluctuations in the applied current and the manufacturing process, particularly during nanofabrication, can directly affect the sensitivity of the proposed magnetometer. The tolerance investigations for the applied currents of 100 and 200 µA yielded similar results, which have been added to the Supplemental Materials (S.5).

## CONCLUSION

This study introduced a novel SOM leveraging BP multilayers, offering a highly sensitive, compact, and energy-efficient alternative to conventional atomic-based magnetometers. By utilizing BP's LD property within a metasurface cavity, the proposed SOM achieves magnetic field detection with sub-nanotesla sensitivity and vector sensing capability. The integration of BP multilayers enhances light-matter interactions, enabling tunable optical responses driven by Lorentz force-induced cavity deformations. The simulation results validate the device's dynamic range adaptability, achieving detection thresholds below 10 nT with an $R^2 > 0.999$ linearity, demonstrating a highly linear response. Additionally, the metasurface cavity was optimized to enhance polarization-dependent absorption, significantly improving the contrast between s- and p-polarized light and further refining the sensor's sensitivity.

Beyond confirming its high sensitivity, the results illustrate how applied current serves as a tunable parameter for optimizing the SOM's performance. By adjusting the current, the trade-off between sensitivity and dynamic range can be controlled, making the device adaptable for different applications. In scenarios requiring ultra-weak field detection, such as biomagnetic sensing in magnetoencephalography (MEG) and magnetocardiography (MCG), higher applied currents enhance sensitivity, achieving detection limits as low as 31.25 pT at 200 µA. Conversely, for broader field detection applications, such as material characterization and industrial sensing, lower currents allow for an expanded dynamic range, reaching up to ±10 nT at 50 µA. Furthermore, the ability to resolve both field magnitude and direction positions this magnetometer as a viable tool for vector field mapping, an essential feature for precision magnetometry in both fundamental research and applied sensing technologies.

This BP-based SOM represents a significant advancement beyond current state-of-the-art magnetic sensors in several critical aspects. Unlike SQUIDs, which require cryogenic cooling at liquid helium temperatures, our device operates at room temperature while still achieving sub-nanotesla sensitivity. Compared to atomic-based OPMs, which are limited to millimeter-scale dimensions due to their reliance on vapor cells, our metasurface design enables miniaturization to the nanoscale—a 2-3 orders of magnitude reduction in size. Furthermore, unlike many solid-state magnetic sensors that require microwatt to milliwatt power levels, our device operates at ultra-low power (≤1 µW), representing an order of magnitude improvement over existing technologies while maintaining comparable or superior sensitivity.

The findings from this work confirm that black phosphorus metasurfaces offer a scalable and high-performance platform for next-generation magnetic field detection. By bridging the gap between atomic-based magnetometers and solid-state nanophotonic technologies, this study establishes BP-based metasurface integration as a transformative approach for ultrasensitive, low-power, and high-resolution magnetic field detection. The demonstrated performance characteristics position this technology for immediate impact in several high-value applications. In neuroscience research, the SOM could enable next-generation neural magnetic field mapping with unprecedented spatial resolution, allowing for more precise brain activity localization than current MEG systems. For wearable health monitoring, the device's small footprint and low power consumption make it ideal for continuous cardiac magnetic field monitoring in compact form factors. In quantum information processing, the high-sensitivity vector detection capabilities could advance spintronic device characterization and quantum state readout. For industrial applications, the sensor could revolutionize non-destructive testing by enabling high-resolution magnetic defect scanning in materials and components. Additionally, its compact size and energy efficiency make it suitable for space-constrained applications like implantable medical devices, autonomous drones, and satellite navigation systems where conventional magnetometers are prohibitively large or power-hungry.